\documentclass{article}

\usepackage{aaai}

\nocopyright

\newcommand{\ra}{\rightarrow}
\newcommand{\la}{\leftarrow}
\newcommand{\lra}{\leftrightarrow}

\newcommand{\Dlogic}{ID-logic}

\newcommand{\eval}[2]{|#2|^{#1}} 
\newcommand{\alphabet}[1]{At_{#1}} 

\newcommand{\rules}[1]{Rules(#1)} 

\newcommand{\Tr}{\mbox{\bf t}}

\newcommand{\II}{{\cal I}}

\newcommand{\ground}[2]{\mbox{$#2$-grounding($#1$)}}

\newcommand{\defp}[1]{Defined(#1)}
\newcommand{\openp}[1]{Open(#1)}

\newcommand{\Def}{{\cal D}}

\newcommand{\clos}[2]{\overline{#2}^{#1}}

\newcommand{\ttt}{\overline{t}}

\newcommand{\xx}{\overline{x}}
\newcommand{\yy}{\overline{y}}
\newcommand{\dd}{\overline{d}}

\newcommand{\body}[1]{
      \begin{array}[t]{l}  #1 \end{array} 
 }

\newcommand{\defin}[2]{
   #1 :: 
  \left \{ 
      \begin{array}{l}  #2 \end{array}  \right \}
 }

\newtheorem{definition}{Definition}

\newtheorem{theorem}{Theorem}

\newtheorem{example}{Example}

\begin{document}

\title{Extending Classical Logic with Inductive Definitions}
\author{ Marc Denecker\\
Department of Computer Science, K.U.Leuven, \\
Celestijnenlaan 200A, B-3001 Heverlee, Belgium. \\
Phone: +32 16 327555 --- Fax:  +32 16 327996 \\
email: marcd@cs.kuleuven.ac.be}
\maketitle

\begin{abstract}
  
  The goal of this paper is to extend classical logic with a
  generalized notion of inductive definition supporting positive and
  negative induction, to investigate the properties of this logic, its
  relationships to other logics in the area of non-monotonic reasoning,
  logic programming and deductive databases, and to show its
  application for knowledge representation by giving a typology of
  definitional knowledge.

\end{abstract}

\section{Introduction}

Since the early days of Computer Science, when John McCarty started to
investigate the role of logic for computing and artificial
intelligence, the key attraction of logic was the promise of a natural
representation of knowledge.
%
%
By the end of the seventies, it had become clear that representing
common sense knowledge in classical logic, despite its clear and
well-understood intuitive reading, was a complex task. Two major
problems were seen as the causes: first, very often common sense
knowledge turns out to be partial and incomplete. Second, in many
applications an unrealistic amount of axioms seemed necessary to
represent the human knowledge; the prototypical example in this
respect is the frame problem \cite{McCarthy69}. To solve these
problems, non-monotonic logics such as circumscription and default
logic, incorporate strong logical closure principles that allow to
reason about partial and incomplete knowledge, and allow a compact
representation of human knowledge.

This paper defines and investigates a conservative extension of
classical logic with a less general non-monotonic closure principle:
the principle of definition and inductive definition.  In the context
of non-monotonic reasoning, definitions have received little attention
so far. This is remarkable for several reasons.  The study of
definitions in knowledge representation has a long tradition which
extends to the origins of Western philosophy. The problem of defining
a {\em natural kind} as studied by the ancient Greek philosophers is
in many ways an instance avant la lettre of the general problem of
common sense knowledge representation.

Also in the context of modern AI, the study of the role of definitions
in common sense knowledge representation has a long tradition. As an
outcome of a series of investigations into the semantics of semantic
networks\footnote{See \cite{Reichgelt91} for a discussion of this
topic.}, Brachman and Levesque \cite{Brachman82} observed that an
important component of expert knowledge is knowledge of the {\em
defining properties} of concepts, and that it is crucial to
distinguish between {\em defining properties} of concepts and {\em
assertional knowledge} on concepts.  Description logics are based on
this idea, and consist of a Tbox to represent definitional knowledge
and an Abox to represent {\em assertional knowledge}. 

Another area related to nonmonotonic reasoning in which definitions
have been prominent is in the context of logic programming and
abductive (or open) logic programming. One of the original ideas on the
declarative semantics of logic programs with negation as failure was
to interpret a logic program as a {\em definition} of its predicates.
This view is underlying Clark's completion semantics \cite{Clark78}.
In \cite{Denecker95a,Denecker95b,VanBelleghem97c}, this view was
extended to abductive logic programs and open logic programs and the
relation with description logics was shown.

In the context of non-monotonic reasoning, Reiter \cite{Reiter96a} and
Amati et al \cite{Amati97a} observed that an important method for
analysis and computation in common sense knowledge representation is
to {\em compile} non-monotonic theories into first order definitions
(i.e. Clark completions). They argue that the advantages of this
compilation are that it clarifies the meaning of the original theory
and that it yields theories that are better suited for computational
purposes. Recently \cite{Denecker98d} and \cite{Ternovskaia98a} showed
a strong correspondence between induction and {\em causality} and
investigated the use of inductive definitions to represent temporal
and causal knowledge.

The above suggests that the study of induction and inductive
definitions could provide a valuable contribution to the area of
common sense knowledge representation, nonmonotonic reasoning and
logic programming. 

The study of induction could be defined as the study of construction
techniques in mathematics. In general, an inductive definition {\em
  defines} a relation (or a collection of relations) through a
constructive process of iterating a recursive {\em recipe} that
defines new instances of the relation in terms of the presence or
absence of other tuples of the same relation or other relations.  The
recipe naturally defines an operator: this operator maps a relation
(or set of relations) to the relation (or set of) that can be obtained
by applying the recipe.

Standard work was done by Moschovakis \cite{Moschovakis74} and Aczel
\cite{Aczel77}. These treatments study the theoretical expressivity of
{\em positive} or {\em monotone induction}.  One spin-off of this
field is fixpoint logic \cite{Gurevitch86}, a subarea of databases
\cite{Abiteboul95}. As shown in \cite{Denecker98c}, the abstract
positive inductive definition logic defined in \cite{Aczel77} is
formally isomorphic with the formalism of propositional Horn programs
under least model semantics.  In the case of Horn programs, the
associated operator corresponds to the immediate consequence operator.

In mathematical applications, also non-monotone forms of induction can
be distinguished:
\begin{itemize}
\item stratified induction and a generalisation, induction in
  well-founded sets. In stratified induction the domain of the defined
  concept(s) can be stratified (possibly in transfinite number of
  levels) such that higher level instances of the concept are defined
  positively or negatively in terms of lower level instances of the
  predicate. Two well-known examples are the following definition of
  even numbers:
\[ \defin{even}{even(0)\la \\even(s(x)) \la \neg even(x)}\]
and the definition of the ordinal powers of a monotone operator as
used in the Tarski-Kleene least fixpoint theory.
  
\item induction in the context of well-founded sets. Here a concept at
  a higher level of the set is defined in terms of the concept at lower
  levels in a monotone or nonmonotone way. An example of definition
  with positive and negative induction in the context of well-founded
  sets is given in \cite{Denecker98c}: the definition of {\em depth}
  or {\em rank} of an element in a well-founded set: it is the least
  ordinal strictly larger than the depths of strictly less elements.
  Also the definition of ordinal powers can be seen as an application
  of this principle.
\end{itemize}

Inflationary fixpoint logic is a well-known extension of fixpoint
logic for nonmonotone operators. However, as argued in
\cite{Denecker98c}, this extension does not match the intuition of
non-monotone induction. For example, the semantics of the nonmonotone
definition of $even$ in inflationary semantics would be the set of all
natural numbers. In general, the inflationary fixpoint is not even a
fixpoint of the original operator, and if it is, it is not
necessarily a minimal fixpoint.

Formalizations of definitions with positive and negative induction
were investigated in the area of Iterated Inductive Definitions
\cite{Feferman70,Feferman81}. As argued in \cite{Denecker98c}, the
intuition of such formalisms is simple and natural and corresponds
with the many notions of stratification in logic programming. Yet, it
is also shown that encoding even simple inductive definitions is
extremely tedious; knowledge representation in such systems is
practically impossible.

\cite{Denecker98c} presents a knowledge theoretic study of a
generalized principle of inductive definition in the abstract setting
of an (infinitary) propositional logic. This definition logic extends
Aczel's infinitary propositional logic for positive
induction\cite{Aczel77}\footnote{Both logics are {\em abstract} in the
sense that infinitary theories and rules are considered.}.  When
extending the notion of inductive definition to unrestricted forms of
induction, a challenging problem arises of defining a uniform
semantical principle that assigns the right semantics to all kinds of
inductive definitions. The main contribution of \cite{Denecker98c} was
to show that the principle of well-founded model in logic programming
\cite{VanGelder91} is the suitable mathematical principle of
generalized inductive definition. The well-founded model is obtained
as the least fixpoint of the 3-valued stable operator
\cite{Przymusinski90b}. The latter operator is a general and robust
implementation of the principle of positive induction; negative
induction is dealt with by iterating this positive induction operator
in a least fixpoint computation. As a consequence, this semantics
generalizes all well-known ways of defining concepts.  It coincides
with least fixpoint and least model semantics in the context of
positive induction; it coincides with iteration of the positive
induction in the context of iterated (or stratified) systems of
inductive definitions; it coincides with Clark completion semantics in
case of non-inductive definitions and inductive definitions on a
well-founded set; beyond these classes, it gives the intended meaning
to mixtures of positive and negative induction in semi-well-founded
sets\footnote{A semi-order $\leq$ is a reflexive, transitive relation
and defines an equivalence relation $x \equiv y$ iff $x \leq y \land y
\leq x$). The set of equivalence classes is a poset. A semi-order is
well-founded if the poset of equivalence classes is well-founded.}.

The present paper is concerned with the role of induction in common
sense knowledge representation and the definition of a knowledge
representation logic based on inductive definitions. To this end, the
abstract logic of \cite{Denecker98c} is lifted to a predicate logic
suitable for representing this generalized notion of inductive
definitions in the context of uncertainty and incomplete knowledge. I
investigate formal properties and methodological guidelines of this
logic, important from the point of view of knowledge representation. A
number of important applications of this logic are sketched.  The
relationship with many other logics are discussed, in particular logic
programming and some of its extensions.

By lack of space, proofs of all theorems are omitted.

\section{An abstract logic of inductive definitions}
\label{SGID}

In \cite{Denecker98c}, I proposed an extension of Aczel's logic for
general monotone and non-monotone inductive definitions. The result is
isomorphic with the formalism of infinitary propositional logic
programs (with negation) under well-founded semantics. An abstract
inductive definition (ID) $D$ in this logic defines a set $\defp{D}$
of symbols, called the set of defined symbols, by a set of rules of
the form
\[ p \la B\]
where $p$ is a defined atom and $B$ a set of positive or negative
literals. The other atoms are called {\em open atoms}; their set is
denoted $\openp{D}$.

In \cite{Denecker98c} it was argued that Przymusinski's 3-valued
extension \cite{Przymusinski90,Przymusinski90b} of Gelfond and
Lifschitz' stable model operator \cite{Gelfond88} is a general and
robust implementation of the principle of positive induction, and that
its least fixpoint, the well-founded model \cite{VanGelder91}
naturally extends the ideas of Iterated Inductive Definitions and
gives the right semantics to generalized inductive definitions.

In general, given an ID $D$ and an interpretation $I$ of the open
symbols of $D$, there is a unique well-founded model extending $I$.
This model will be denoted $\clos{D}{I}$.  An interpretation is a
model of $D$ iff $M = \clos{D}{M_o}$ where $M_o$ is the restriction of
$M$ to the open symbols.  In general, a model of an inductive
definition is a partial (3-valued) interpretation. However, for broad
classes of definitions, the well-founded model is known to be total
(2-valued). These are the cases that are of interest in this paper.

\section{\Dlogic: classical logic with definitions} 
\label{SCLGID} 

This section defines a conservative extension of classical logic with
definitions. An \Dlogic\ theory $T$ (based on some logical alphabet
$\Sigma$) consists of a set of classical logic sentences and a set of
definitions. A definition $D$ is a pair of a set $\defp{D}$ of 
predicates and a set $\rules{D}$ of rules of the form:
\[ p(\ttt) \la F\]
where $p \in \defp{D}$ and $F$ an arbitrary first order formula based
on $\Sigma$. Predicates of $\defp{D}$ are called the {\em defined
  predicates} of $D$; other predicates are called {\em open
  predicates} of $D$.  A definition defines the defined predicates in
terms of the open predicates. More precisely, given some state of the
open predicates, the rule set of the definition gives an exhaustive
enumeration of the cases in which the defined predicates are true; any
defined atom not covered by a rule is defined as false.

A definition will be formally represented as in the example:
\[ \defin{even,odd}{ 
        even(0) \la \\ 
        even(S(x)) \la odd(x) \\ 
        odd(S(x)) \la even (x)  }
\]
This is one definition defining two predicates simultaneously.

In \Dlogic, definitions are considered as sentences. An \Dlogic\ 
theory based on $\Sigma$ consists of sentences and may contain
different definitions, even for the same predicates. An
$\Sigma$-interpretation is a model of an \Dlogic\ theory iff it is a
model of all its sentences. So, it suffices to define what is a model
of a definition.

The semantics for propositional definitions of section \ref{SGID} can
be lifted  easily to the predicate case by use of the grounding
technique: the technique of reducing a predicate definition to an
infinitary propositional definition\footnote{In \cite{VanGelder93}, an
  alternative way of defining the well-founded semantics of predicate
  rules is proposed; it is based on a different treatment of positive
  and negative occurrences of predicates in the body of rules. I
  believe both techniques are equivalent but haven't proven this. }.
In the context of \Dlogic, this grounding of a definition is
constructed using the domain and functions of some (general
non-Herbrand) interpretation $I$.

To define the grounding the following terminology is needed. Given an
alphabet $\Sigma$ and a $\Sigma$-interpretation $I$, define the
alphabet $\Sigma_I$ by adding the domain elements of $I$ as constants
to $\Sigma$\footnote{Note that $\Sigma_I$ may be infinite, even
non-countable. This is mathematically and philosophically
non-problematic because $\Sigma_I$ is purely used as a semantic
device, namely to define the grounding.}. $I$ is naturally extended to
$\Sigma_I$ by defining $I(x)=x$ for each domain element $x$ of $I$.
The evaluation of a ground term $t$ of $\Sigma_I$ (which may contain
domain elements of $I$) is defined inductively as usual, and is
denoted $\eval{I}{t}$. Likewise, truth value of a sentence of
$\Sigma_I$ is defined by the usual truth recursion.

Given some partial (3-valued) interpretation $I$ and a definition $D$,
$I_o$ denotes the restriction of $I$ to the constant, functor and open
predicate symbols of $D$. $\alphabet{I}$ denotes the set of all atoms
$p(\dd)$ where $p$ is a defined predicate of $D$ and $\dd$ is a tuple
of domain elements of $I$. A ground instance of a rule $p(\ttt[\xx])
\la F[\xx]$ with $\xx$ the tuple of all its free variables, is a rule
$p(\ttt[\dd]) \la F[\dd]$ obtained by substituting domain elements
$\dd$ for $\xx$. 

The intuition to get the grounding is as follows. We must replace a
rule-instance $p(\ttt[\dd]) \la F[\dd]$ by some logically equivalent
set of propositional rules $A \la B$ where $B$ is a set of
literals. We will select $A = p(\eval{I}{\ttt[\dd]})$ and $B$ any set
of literals of $\alphabet{I}$ such that if all elements of $B$ were
true, then $F[\dd]$ would be true. Or, the bodies of the grounding
correspond to the partial models of the rule body $F[\dd]$.

Define a consistent set $S$ of literals of $\alphabet{I}$ as a set of
positive or negative literals based on $\alphabet{I}$ that does not
contain a pair of complementary literals $p(\dd), \neg p(\dd)$. Note
that there is a one-to-one correspondence between partial
interpretations extending $I_o$ and consistent sets of
$\alphabet{I}$-literals.  Each partial interpretation $J$ extending
$I_o$ defines a unique consistent set $S_J$ of all literals $l$ of
$\alphabet{I}$ that are true in $J$. Vice versa, each consistent set
$S$ defines a unique partial interpretation $J_S$ extending $I_o$ such
that $J_S(l)=\Tr$ iff $l\in S$. Moreover, $J_{S_J}=J$.

\begin{definition}\label{Dgrounding}
Given an interpretation $I$, the grounding of a definition $D$
w.r.t. $I$, denoted $\ground{D}{I}$, is the propositional definition
defining all atoms of $\alphabet{I}$ and consisting of all rules
\[  p(\eval{I}{\ttt[\dd]}) \la S_J \]
such that $p(\ttt[\dd]) \la F[\dd]$ is a ground instance of a rule of
$D$ and $J$ is a partial model of $F[\dd]$ extending $I_o$.
\end{definition}

\begin{definition}\label{Dmodel}
A 3-valued interpretation $I$ is a justified interpretation of $D$ iff
$S_I$ is the 3-valued (well-founded) model of the grounding of $D$
w.r.t. $I$. $I$ is a justified interpretation of a theory $T$ iff it
is a justified interpretation of all its definitions and a (3-valued)
model of the classical logic sentences of $T$.

An interpretation $I$ is a model of a definition $D$, resp. theory
$T$, iff it is a total (i.e. 2-valued) justified interpretation of
$D$, resp. $T$.
\end{definition}

The above model theory is based on total, general non-Herbrand models.
As a consequence, \Dlogic\ is an extension of classical logic. The
restriction to total models is not only necessary to get a extension
of classical logic, but also because of methodological constraints on
the use of definitions, as explained in the next section.

\begin{example}
  The first example shows that different definitions are independent
  and interact in a monotonic way. Consider the theory consisting of
  three definitions.
\[ \left \{\begin{array}{l}
\defin{father}{ father(x,y) \la \body{parent(x,y)\land\\ male(x)}}\\
\defin{mother}{ mother(x,y) \la \body{parent(x,y)\land\\ female(x)}}\\
\defin{parent}{ parent(x,y) \la father(x,y)\\
                parent(x,y) \la mother(x,y)} 
\end{array} \right \}
\]
Note that in the first definition, $father$ depends on $parent$, while
in the third, $parent$ depends on $father$. However, the semantics of
a set of definitions is monotonically composed of the semantics of
its definitions. Since none of these definitions is recursive, each is
equivalent with its completed definition.  Consequently, this triple
of definitions is equivalent with the FOL theory:
\[ \left \{\begin{array}{l}
father(x,y) \lra parent(x,y)\land male(x)\\
mother(x,y) \lra parent(x,y)\land female(x)\\
parent(x,y) \lra father(x,y) \lor mother(x,y)
\end{array} \right \}
\]
One can observe that if $male(x) \lra \neg female(x)$ holds, then the
definition of $parent$ is redundant.

Compare this theory with the simultaneous definition obtained by
merging the three definitions in one:
\[ \left \{ \begin{array}{l}
father, mother,parent::\\
\left\{ \begin{array}{l}
  father(x,y) \la parent(x,y)\land male(x)\\
  mother(x,y) \la parent(x,y)\land female(x)\\
  parent(x,y) \la father(x,y)\\
  parent(x,y) \la mother(x,y) \end{array}  \right \} \end{array} \right \}
\]
This new definition is positive recursive. This has the unintended
effect that in each model, $father$, $mother$ and $parent$ are interpreted
as the empty relationships.
\end{example}
\begin{example}
A theory may contain more than one definition for the same concept. E.g.
\[ \left\{\begin{array}{l} 
    \defin{even}{even(0) \la \neg odd(x)}\\
    \defin{even}{even(0)\\
                 even(s(s(x))) \la even(x)}\end{array} \right\}
\]
The first definition defines even as the complement of odd. In itself,
it does not define $even $ nor $odd$. The second definition defines
$even$ by the usual positive recursion. The combination of both definitions
defines both $even$ and $odd$. 
\end{example}

\begin{definition}\label{Dspecialdef}
  A definition is recursive iff a defined predicate appears in the
  body of a rule. A definition is positive recursive iff all
  occurrences of the defined predicates in the body of the rules are
  positive (i.e. occur in the scope of an even number of negations). A
  simultaneous definition defines more than one predicate. A
  stratified definition is one in which the defined predicates can be
  semi-ordered such that each defined predicate occurring positively,
  resp. negatively, in the body of a rule is less, resp.  strictly
  less, than the predicate in the head.
  
  A definition hierarchy is a set ${\cal D}$ of definitions such that each
  predicate is defined in at most one definition of ${\cal D}$ and ${\cal D}$
  can be ordered such that each open predicate appearing in a
  definition is not defined in a later definition.
\end{definition}
 
Below, I define the concept of a well-founded definition. This concept
generalizes the principle of definition in a well-founded set.
\begin{definition}\label{Dwellfounded}
  A definition $D$ is well-founded in some collection $\II$ of total
  interpretations of the open predicates of $D$ iff for each $I \in
  \II$, there exists a well-founded order on the atoms of
  $\alphabet{I}$ such that for each ground instance $p(\ttt[\dd]) \la
  F[\dd]$, the body $F[\dd]$ has the same truth value in all partial
  interpretations that extend $I$ and are identical on all atoms less
  than $p(\eval{I}{\ttt[\dd]})$.
\end{definition}

The following theorem states an interesting property of well-founded
definitions.
\begin{theorem}\label{TWellfounded}
  If $D$ is well-founded in $\II$, then each justified interpretation
  $M$ of $D$ extending an element $I$ of $\II$ is total (and hence a
  model) and coincides with the least model of the 3-valued completion
  of $D$ \cite{Fitting85} extending $I$. $M$ is the unique model of the
  Clark completion \cite{Clark78} of $D$ extending $I$.
\end{theorem}

\begin{example}\label{EWellfounded}
  Consider the definition of even numbers:
\[
\defin{even}{ even(0) \la \\ even(S(x)) \la \neg even(x)}
\]
In the context of the natural numbers, this definition is well-founded
and the justified interpretation is total. However, in any
interpretation where the successor function contains cycles, the
justified interpretation is partial.
\end{example}

\section{Properties of definitions}
\label{SProp} 

\subsection{Consistency of definitions}

The aim of an inductive definition is to {\em define} its defined
predicates. Therefore, a natural quality requirement is that those
justified interpretations that are total in the open predicates,
should define truth of all defined predicates, i.e. they should be
total in all predicates. As shown by Example \ref{EWellfounded} the
property of having total justified interpretations is context
dependent.

\begin{definition}\label{Dwelldefining}
  A definition $\Def$ is well-defining in a collection $\II$ of total
  interpretations of its open predicates iff each justified
  interpretation of $\Def$ extending an element of $\II$ is total.
  Otherwise, $\Def$ is called an unfounded definition in $\II$.
\end{definition}

Part of the knowledge representation methodology for representing
definitions is to show that each definition in the theory is
well-defining in the collection of relevant interpretations of its
open predicates. For this purpose, practical mathematical techniques
must be developed.

\begin{theorem}
  Assume that a theory $T$ can be split up in a sequence of theories
  $T_1,..,T_n$ such that for each $i$, the predicates with a
  definition in $T_i$ do not appear in $T_1,..,T_{i-1}$ and for each
  model $I$ of $T_1\cup..\cup T_{i-1}$, the definitions in $T_i$ are
  well-defining in $I$.
  
  Then each justified interpretation of $T$, total for the subset of
  predicates without definition in $T$, is total.
\end{theorem}
The proof of this theorem is omitted.

Some syntactic properties that guarantee that a definition is
well-defining in every context are well-known from the logic
programming literature:
\begin{itemize}
\item non-recursive definitions
\item positive recursive definitions
\item stratified  definitions
\end{itemize}

Other properties guarantee well-defining definitions in some specific
context. Inductive definitions corresponding to acyclic \cite{Apt90}
or locally stratified logic programs \cite{Przymusinski88} are
well-defining in the context of Herbrand interpretations. It follows 
from theorem \ref{TWellfounded} that a well-founded
definition in context $\II$ is  also well-defining in $\II$. 

A syntactical criterion that guarantees well-foundedness and hence
well-defining-ness is the following. Define a relativized definition
w.r.t. some strict order $<$ as a definition of a predicate
$p(x,\yy)$ that consists of rules:
\[ p(x,\ttt) \la F[x] \]
such that each $p$-atom in $F$ is of the form $p(z,\ttt')$ and appears
in the scope of a subformula of $F[x]$ of the form $\forall z. z<x \ra
G$ or $\exists z. z<x \land G$.

When $<$ represents a well-founded order, relativized definitions are
well-founded and hence well-defining, by theorem \ref{TWellfounded}.
The following theorem was proven.
\begin{theorem}
  A relativized definition (w.r.t. to $<$) is well-founded in each
  interpretation that interprets $<$ as a strict well-founded order.
\end{theorem}

\subsection{Equivalence of definitions}

In a logic for knowledge representation, there should be a
well-understood notion of equivalence. The following example shows
that one cannot simply replace bodies of cases by equivalent bodies
(w.r.t. 2-valued semantics).
\begin{example}
  The definitions $\defin{p}{ p \la \Tr}$ and $\defin{p}{ p \la p \lor
    \neg p}$ have different justified interpretations, respectively
    the interpretations (represented as literal sets) $\{ p\}$ and
    $\{\}$. Note that their bodies are equivalent w.r.t. 2-valued
    semantics but not w.r.t.  3-valued semantics.
\end{example}

Some important cases of equivalence preserving rules are sketched
below:
\begin{itemize}
\item A case $p(\ttt[\xx]) \la F$ can be replaced by $p(\yy) \la
  \exists \xx. \yy=\ttt[\xx] \land F$.
\item In a definition, two cases $p(\ttt) \la F_1$ and $p(\ttt) \la
  F_2$ can be replaced by one case $p(\ttt) \la F_1 \lor F_2$.
  Together with the first rule, it follows that replacing a set of rules
  by its Clark completion is equivalence preserving.
\item The substitution of a sub-formula $F[\xx]$ in the body of a case
  of a formula by a formula $G[\xx]$ is equivalence preserving if
  $F[\xx]$ and $G[\xx]$ are equivalent in 3-valued logic, i.e. if
  $\forall \xx. F[\xx] \lra G[\xx]$ is a tautology in 3-valued
  logic\footnote{Here the strong Kleene truth table for $\lra$ must be
    used.}.
\item Define the composition of two definitions $\defin{Pred_1}{C_1}$
  and $\defin{Pred_2}{C_2}$ as the definition $\defin{Pred_1\cup
    Pred_2}{C_1\cup C_2}$. In general, substituting a pair of
  definitions by their composition is not equivalence preserving.
  \cite{Verbaeten99a} presents an extensive study of when merging
  definitions is equivalence preserving in the context of open logic
  programming, a sub-formalism of the logic defined here. One
  important example is that a definition hierarchy (Definition
  \ref{Dspecialdef}) is equivalent with its composition. Note that the
  composition of a definition hierarchy of positive recursive
  definitions is a stratified definition.
\end{itemize}

\subsection{Monotonicity, non-monotonicity and modularity}

Non-monotonicity is a necessary property of elaboration tolerant logic
descriptions \cite{McCarthy98}. Non-monotonicity is a natural
consequence of any sort of closure principle. However, also a degree
of monotonicity is important in knowledge representation. Indeed, a
highly desirable feature of a knowledge representation logic is that
independent properties of the problem domain can be represented in a
modular way, and that adding the modules together in one theory
preserves the semantics of each module. Modular composition is
guaranteed if the models of the composition of the modules are models
of the independent modules. However, this property guarantees also
that the extension of one module with another is a monotonic
operation.

Both properties are present in the logic defined here: 

\begin{itemize}
\item Extending a definition with one or more new cases is in general a
  non-monotonic operation.
\item Adding new definitions or new axioms to a theory is a monotonic
  operation. This follows trivially from the definition of model.
\end{itemize}


\section{Applications of definitions}
\label{STypology}

Below some applications of ID-logic are given.
\begin{description}
\item {\bf Terminological knowledge.}  
  
  According to \cite{Brachman82}, definitions of terminology
  constitutes an important part of expert knowledge. Terminological
  knowledge is about the {\em defining properties} of a concept, i.e.
  the necessary and sufficient conditions to belong to the concept.
  To see the crucial difference between terminological and assertional
  knowledge, consider the atomic statement $c(O)$ that some object $O$
  belongs to a concept $c$. When $c(O)$ represents an assertional
  statement, it asserts that $O$ belongs to $c$ and, hence, satisfies
  the defining property of $c$. On the other hand, if the atom is
  added as a new {\em case} to the definition of $c$, then the
  defining property of $c$ is modified such that $O$ belongs to $c$ by
  definition and not by virtue of its properties.\vspace{2mm}
  
\item {\bf Temporal Reasoning.}
  
  In \cite{Ternovskaia98a}, it was shown that Reiter's situation
  calculus \cite{Reiter91} has an equivalent formalization by a set of
  positive recursive definitions of the fluent predicates and of the
  effects of actions. Using general inductive definitions (with
  positive and negative induction), the formalization can be further
  simplified in \Dlogic. Below, I sketch how to do this. 
  
  The definition defines all fluent symbols and all causal predicates
  by simultaneous induction on the poset of situations. I introduce
  for each fluent $f$ three new predicates: $initially_f$ to represent
  the initial state of $f$, and $cause_f$ and $cause_{\neg f}$,
  representing initiating and terminating causes for $f$.  For each
  fluent symbol $f$, the definition contains three cases\footnote{We
  assume a many-sorted version of \Dlogic, with situation, action and
  user defined sorts.}:
\[ \begin{array}{l}
f(\xx,S_0) \la initially_f(\xx)\\
f(\xx,do(a,s)) \la cause_f(a,s,\xx)\\
f(\xx,do(a,s)) \la f(\xx,s) \land \neg cause_{\neg f}(a,s,\xx) 
\end{array}\]

Note that in contrast to Reiter's situation calculus, this rule set
does not contain a rule of the form:
\[ \neg f(\xx,do(a,s)) \la cause_{\neg f}(a,s,\xx)\]
However, it is easy to show that the completion of the above 3 rules
entails the formula:
\[  \neg f(do(a,s)) \la \neg cause_f(a,s,\xx) \land cause_{\neg f}(a,s,\xx)\]
which reduces to the causal rule for $\neg f$ if the natural
requirement is added that an action cannot cause $f$ and $\neg f$ in
the same situation. This requirement is formalised by the clause:
\[ \la cause_{\neg f}(a,s,\xx), cause_f(a,s,\xx)\]
This illustrates a general methodological principle of using inductive
definitions. In an inductive definition, one defines a concept by
enumerating the positive cases; given such an enumeration, the closure
mechanism of the semantics yields the negative cases.

In addition, per initiating effect of some action represented by an
action term $A[\yy]$, there is a case:
\[ cause_f(A[\yy],s,\xx) \la \Psi[\yy,s,\xx]\]
such that the only term in $\Psi$ of the situation sort is $s$ and it
appears purely in fluent symbols.  Likewise, for each terminating
effect there is a case:
\[ cause_{\neg f}(A[\yy],s,\xx) \la \Psi[\yy,s,\xx]\]

\begin{theorem}
  A definition consisting of the above rules is well-founded in the
  collection of all interpretations that satisfy the Unique Names
  Axioms (UNA) axioms \cite{Reiter80a} and the second order induction
  axiom for the situation sort\footnote{The order of the atoms, needed
  to establish the well-foundedness of the definition is the order
  generated by the atoms $f(..,do(a,s)) > cause_f(a,s,..) ,
  cause_{\neg f}(a,s,..) > g(..,s)$, with $f$, $g$ arbitrary
  fluents.}.
\end{theorem} 
It follows from theorem \ref{TWellfounded} that the semantics of this
inductive definition coincides with its Clark completion. Note that
the completion of this definition is very similar to Reiter's state
successor axioms \footnote{The main difference is that it lacks the
action precondition predicate $poss(a,s)$. Action preconditions can be
added to the inductive definition, by extending the definition with
cases defining $poss$, by adding the atom $poss(a,s)$ as a conjunct to
the second and third case defining $f$ and adding a fourth rule
\[f(\xx,do(a,s)) \la f_o(\xx,do(a,s)) \land
\neg poss(a,s)\] where $f_o$ is a new open predicate.}.

The inductive definition representation of situation calculus in
\Dlogic\ represents initiating and terminating effects in a
case-by-case way. This results in a modular, elaboration tolerant
representation of the domain in the sense that one can easily add new
cases or drop or refine existing ones. 
This definition can be further extended with definitions for defined
fluents, e.g. the definition of the transitive closure of physical
connections in a computer network, in the context in which these
physical connections may change:
\[ \begin{array}{l} 
  connected(c1,c2,s) \la physical\_connection(c1,c2,s) \\
  connected(c1,c2,s) \la \body{connected(c1,c3,s) \land \\ connected(c3,c2,s)}
\end{array} 
\]
Also, similarly as in \cite{Ternovskaia98a}, ramification rules can be
added to this theory.\vspace{2mm}

\item {\bf Inductive definitions as an approach to Causality.}

In \cite{Denecker98d} we argued that inductive definitions are a
suitable formalization of causality. Causality information is an
example of constructive information. Effects and forces propagate in a
dynamic system through a constructive process in the following sense:
\begin{itemize}

 \item there are no deus ex machina effects. Each effect has a cause;
 it is caused by a nonempty combination of actions and other effects.
 
 \item The causation order among effects is a well-ordering. I.e.
 there is no pair of effects each of which have caused the other;
 stronger, there is no infinite descending chain of effects each of
 which has been caused by the previous one in the chain.

 \end{itemize}

The construction process of an inductive definition formally mimics
this physical process of the propagation of the causes and
effects. Based on this idea, \cite{Denecker98d} proposes a general
solution to model ramifications. One point of \cite{Denecker98d} was
that effects may easily depend on both presence and absence of other
effects. For example, in the case that one latch of a suitcase is
open, the effect of opening the second latch produces a derived effect
of opening the suitcase, but only if the first latch is not closed
simultaneously. As a consequence, if fluents mutually can influence
each other, descriptions of ramifications may easily contain positive
and negative loops. As shown in \cite{Denecker98d},  the
well-founded semantics deals well with these loops.\vspace{2mm}

\item {\bf Induction axioms; Domain Closure Axiom (DCA).}

As a conservative extension of classical logic, \Dlogic\ assumes
uncertainty on the domain of discourse (due to the non-Herbrand
interpretations). The Domain Closure Axiom (DCA) expresses that the
domain of discourse contains only named objects.  In
\cite{McCarthy80}, McCarthy showed how the DCA can be represented by a
combination of circumscription on a set of rules and a FOL
assertion. The mapping to \Dlogic\ is straightforward. The set of
rules is an inductive definition of a new predicate $U$; it consists
of one case per constant $C$ and per functor $f$:
\[ \defin{U}{ U(C) \la \\ .. \\ U(f(\xx)) \la U(x_1),..,U(x_n) } \] This
defines $U$ as the set of all named objects. The FOL axiom expresses
that all objects in the domain are named\footnote{Note here the
distinction between defining knowledge and assertional knowledge. If
one would add the FOL assertion as a case to the inductive definition,
then $U$ would be defined to be the complete domain of discourse.}:
\[ \forall x. U(x) \]
The DCA is a generalized induction axiom. In the case of the language
of the natural numbers (0 and $S/1$), the above formalization of the
DCA is equivalent with Peano's second order induction axiom. The
induction axiom for situations as needed in Reiter's situation
calculus can be expressed in a similar way.

The semantics of many logics, e.g. logic programming and deductive
databases, is based on Herbrand interpretations.  This introduces the
implicit ontological constraint that all terms in the domain
of discourse are named. This constraint is absent in classical logic
and in \Dlogic\ but can be explicitly formalized by the pair of the
DCA and the Unique Names Axioms (UNA) \cite{Reiter80a} or the Clark
Equality Theory (CET) \cite{Clark78}. It is easy to show that each
model of  DCA+UNA is isomorphic with a Herbrand interpretation.\vspace{2mm}

\item {\bf Tables.} 

The simplest way of defining a concept is by exhaustive enumeration of
its elements. A table, as in the context of databases, can naturally
be viewed as a definition by exhaustive enumeration. Tables are
commonly used to define concepts, not only in databases but also in
common sense knowledge representation, e.g. to define some scenario.
\end{description}

\section{Relationships to other logics.}
\label{SRelations}

\Dlogic\ shows tight relationships with a class of different logics in
different areas of AI, computer science and mathematical
logic. Earlier in this paper, strong relations to circumscription and
Clark completion came to light. Despite differences in the syntactic
sugar, there are strong relationships between description logics and
\Dlogic. \Dlogic\ fits into the schema of description logics, with a
Tbox consisting of the definitions and an Abox consisting of classical
logic axioms. The correspondence between description logics and a
sub-logic of \Dlogic\ were formally investigated in
\cite{VanBelleghem97c}.

Also fixpoint logic can be embedded in \Dlogic, modulo syntactic
sugar. The difference in syntactic sugar is that  in \Dlogic,
the defined concepts are named by global predicate symbols, whereas in
fixpoint logic, the defined concepts are represented by an operator
form\footnote{E.g. the transitive closure of a predicate $p$ is
represented by the form:
\[\Gamma_{\Psi}^{(x,y)}(p(x,y) \lor (\exists z. \Psi(x,z) \land \Psi(z,y)))\] 
It corresponds to the \Dlogic\ definition of a binary predicate $tr$
defined by:
\[\defin{tr}{tr(x,y) \la p(x,y) \lor tr(x,z) \land tr(z,y)}\]
}. Inflationary fixpoint logic is an extension of
fixpoint logic for fixpoint forms with negation in it. However, as
mentioned in the introduction, this logic doesn't give the intended
semantics to inductive definitions with negation. For example, the
concept of even numbers is represented in \Dlogic\ by:
\[\defin{even}{even(x) \la x=0 \lor \exists y. x=S(y) \land \neg
even(y)}\] 
The corresponding inflationary fixpoint form
\[\Gamma_{\Psi}^{x}(x=0 \lor \exists y. x=S(y) \land \neg \Psi(y))\] 
denotes the set of all natural numbers. 

Logic Programming can be embedded in ID-logic in a
straightforward way. Some of its extensions can be embedded as well.
Abductive logic programming \cite{Kakas93a} (or open logic programming,
as it is called in \cite{Denecker95a}) can be embedded also in
ID-logic. An abductive logic framework is a triple $< A,P,T>$ of a set
$A$ of abducible predicates, a set $P$ of rules defining non-abducible
predicates and a set $T$ of FOL axioms, called {\em constraints}. Its
embedding in ID-logic is trivial: it is the theory $T\cup \{D_p\}$
where $D_P$ is a definition with $\rules{D}=P$ and with $\defp{D}$ the
set of non-abducible predicates.  In this embedding, the semantics of
an abductive logic program is given by general non-Herbrand
well-founded models.

The formalism of deductive databases \cite{Abiteboul95} descends from
logic programming.  A deductive database consists of a triple
(EDB,IDB,IC) where EDB (extensional database) consists of tables for
extensional predicates, IDB (intensional database) consists of a logic
program defining intensional predicates in terms of intensional and
extensional predicates, and IC (integrity constraints) is a set of FOL
axioms. The embedding of a deductive database in \Dlogic\ consists of
a set of definitions (by exhaustive enumeration) for the extensional
predicates, a simultaneous definition for the intensional predicates,
the set IC as FOL axioms and finally, the DCA+UNA.

\section{\Dlogic\ compared to NMR}
\label{SComparisonNMR}

As argued, definitional knowledge is an important component of human
expert knowledge. Such knowledge consists of definitions of
terminology but is not restricted to that: e.g. induction axioms,
knowledge of physical causation, etc..  \Dlogic\ is specifically
designed for representing such knowledge.

With respect to other applications such as the representation of
defaults, the scope and applicability of inductive definitions is more
restricted than other non-monotonic principles such as circumscription
and default logic. The use of definitions for such forms of common
sense knowledge requires a more rigorous methodology of a priori
analysis and restructuring of the knowledge on the problem domain. 

As mentioned earlier, Reiter \cite{Reiter96a} and Amati et al
\cite{Amati97a} observe that {\em compilation} of non-monotonic
theories into first order definitions (in Beth's style) can produce
theories that clarify the meaning of the original theories and are
computationally more attractive.  Using the more general notion of
inductive definitions in \Dlogic, this compilation of non-monotonic
theories to definitions becomes easier and can be performed for larger
classes of NMR theories. Moreover, due to the non-monotonicity of the
logic, the compiled representation will often maintain most of the
elaboration tolerance of the original NMR representation.

With respect to computational efficiency, reasoning on \Dlogic\ is
undecidable in general, like in the case of other predicate
non-monotonic logics. However, there is plenty of evidence that in many
cases, definitions can be reasoned on more efficiently that other
non-monotonic formalisms or even classical logic (e.g. constructing a
well-founded model of a propositional definition is polynomial, while
constructing a model of a propositional theory is NP-complete).  A
common aspect of virtually all logics related to \Dlogic, is the
strong focus on efficient implementation. Techniques from these areas
can be used to implement efficient solvers.


\section*{Acknowledgements}
This work has benefitted from discussions with the following people:
Maurice Bruynooghe, Danny Deschreye, Michael Gelfond, Vladimir
Lifschitz, Victor Marek, Ray Reiter, Eugenia Ternovskaia, Mirek
Truszczynski, Kristof Van Belleghem. Thanks!

\end{document}